# Bifurcation and control of chaos in Induction motor drives

Krishnendu Chakrabarty
Kalyani Government Engineering College
Kalyani-741235, India
chakrabarty40@rediffmail.com

Urmila Kar
National Institute of Technical Teachers' Training and Research
Kolkata-700106, India
urmilakar@rediffmail.com

***Abstract*— Induction motor controlled by Indirect Field Oriented Control (IFOC) is known to have high performance and better stability. This paper reports the dynamical behavior of an indirect field oriented control (IFOC) induction motor drive in the light of bifurcation theory. The speed of high performance induction motor drive is controlled by IFOC method. The knowledge of qualitative change of the behavior of the motor such as equilibrium points, limit cycles and chaos with the change of motor parameters and load torque are essential for proper control of the motor. This paper provides a numerical approach to understand better the dynamical behavior of an indirect field oriented control of a current-fed induction motor. The focus is on bifurcation analysis of the IFOC motor, with a particular emphasis on the change that affects the dynamics and stability under small variations of Proportional Integral controller (PI) parameters, load torque and k, the ratio of the rotor time constant and its estimate etc. Bifurcation diagrams are computed. This paper also attempts to discuss various types of the transition to chaos in the induction motor. The results of the obtained bifurcation simulations give useful guidelines for adjusting both motor model and PI controller parameters. It is also important to ensure desired operation of the motor when the motor shows chaotic behavior. Infinite numbers of unstable periodic orbits are embedded in a chaotic attractor. Any unstable periodic orbit can be stabilized by proper control algorithm. The delayed feedback control method to control chaos has been implemented in this system.**

## I. INTRODUCTION

Nowadays, induction motors are extensively used for household and industrial applications for its relatively rugged and inexpensive structure. Therefore, much attention is given to design controllers to control the speed of the motor for various applications. The vector control of AC drives has been used widely in high performance control. Indirect field oriented control (IFOC) is one of the most effective vector control method to control induction motor due to the simplicity of design, better stability and high performance.

The effect of error in estimation of rotor resistance of IFOC motor in the light of stability has been investigated [1]. It has been previously shown that the IFOC based speed control of induction motors with constant speed is globally stable [2]. An analysis of saddle-node and Hopf bifurcation in

IFOC drives due to errors in the estimate of the rotor time constant provides a guideline for setting the gains of PI speed controller in order to avoid Hopf bifurcation [3].Stable drives are required for conventional application. Hopf bifurcation in IFOC drives with estimation error of rotor time constant as bifurcation parameter and condition of Hopf bifurcation are analyzed [4]. Bifurcation causes instability in IFOC motors. Researchers investigated only Hopf bifurcation in IFOC motor. But there are existences of other types of bifurcation in these motors. Exhaustive studies of bifurcation phenomenon of IFOC motors are not available in the literature. An attempt is made to fill this gap in this paper.

The strange attractor of a chaotic system is a collection of infinite periodic orbits embedded in the system's phase plane. All these periodic orbits are in the unstable state. As for example, the trajectory in the chaotic attractor comes very near to period-1 orbit and goes away with time without becoming stable period-1 orbit. The same is observed for other period-P orbits. This characteristic of chaotic system makes it possible to stabilize any unstable periodic orbit with application of tiny change of parameters or by introduction of control algorithm. In recent years, investigators have shown that by perturbing a chaotic system in right way, one can force the system to follow one of its many unstable behaviors. With proper control, one can rapidly switch among many different behaviors. This gives a clue to improve the response as well as the domain of operation in systems that exhibit chaos for some parameter values.

Various methods to control chaos in different chaotic systems have been developed. An overview of different approaches to the control of chaos for various nonlinear dynamic systems has been reported by G.Chen el al [5]. Hence it is possible to ensure desired behavior of the motor even if the behavior of the motor is chaotic due to nonlinearity present in the system and values of parameters of the motor chosen.

Then, study of bifurcation, chaos and control of chaos in IFOC induction motor drive are the key points of investigation of this paper

The paper has been organized as follows. The mathematical model of IM system with IFOC is obtained firstly in section II. The dynamics of the system is given in section III. The detailed bifurcation behaviors of IFOC motor with different parameters are given in section IV. The method to control of chaos is given in section V. The conclusion is given in section VI.

## II. MATHEMATICAL MODEL OF THE SYSTEM

The nonlinear dynamical model of a current fed induction machine is selected for bifurcation analysis. Stator currents are the input to the system. Rotor equations of the induction motor expressed in a reference frame rotating at synchronous speed are given as follows:

$$
\begin{aligned}
&R_r i_{dr} - \omega_{sl}\psi_{qr} + \dot{\psi}_{dr} = 0 \\
&R_r i_{qr} + \omega_{sl}\psi_{dr} + \dot{\psi}_{qr} = 0 \\
&\psi_{qr} = L_r i_{qr} + L_m i_{qs} \\
&\psi_{dr} = L_r + L_m i_{ds}
\end{aligned} \tag{1}
$$

If the machine has distributed windings and sinusoidal MMF, then only d-q variables are taken in torque equation. The electrical torque are expressed as

$$T_e = \frac{3}{2} P \frac{L_m}{L_r}(\psi_{dr} i_{qs} - \psi_{qr} i_{ds}) \tag{2}$$

The mechanical equation is given by

$$\dot{\omega} = \frac{1}{J}(T_e - T_l - \beta\omega) \tag{3}$$

From equation (1), (2) and (3), the simplified model of the induction motor can be expressed as:

$$
\begin{aligned}
&\dot{\psi}_{qr} = -\frac{R_r}{L_r}\psi_{qr} - \omega_{sl}\psi_{dr} + \frac{L_m}{L_r} R_r i_{qs} \\
&\dot{\psi}_{dr} = -\frac{R_r}{L_r}\psi_{dr} + \omega_{sl}\psi_{qr} + \frac{L_m}{L_r} R_r i_{ds} \\
&\dot{\omega} = -\frac{\beta}{J}\omega + \frac{1}{J}\left[\frac{3}{2}\frac{L_m}{L_r} P(\psi_{dr} i_{qs} - \psi_{qr} i_{ds}) - T_l\right]
\end{aligned} \tag{4}
$$

In order to form the state-space equation, the following states, inputs and constants are defined:

$$
\begin{aligned}
&x_1 = \psi_{qr},\ x_2 = \psi_{dr},\ u_1 = \omega_{sl},\ u_2 = i_{ds},\ u_3 = i_{qs} \\
&c_1 = \frac{R_r}{L_r},\ c_2 = \frac{L_m}{L_r} R_r,\ c_3 = \frac{\beta}{J},\ c_4 = \frac{1}{J}, \\
&c_5 = \frac{3}{2}\frac{L_m}{L_r} P
\end{aligned} \tag{5}
$$

Including (5) into (4), the state –space model can be written as :

$$
\begin{aligned}
&\dot{x}_1 = -c_1 x_1 + c_2 u_3 - u_1 x_2 \\
&\dot{x}_2 = -c_1 x_2 + c_2 u_2 + u_1 x_1 \\
&\dot{\omega} = -c_3\omega + c_4\left[c_5(x_2 u_3 - x_1 u_2) - T_l\right]
\end{aligned} \tag{6}
$$

Amongst the constants defined in (5), $c_1$ represents the inverse of the rotor time constants for the fundamental frequency and it has influence on the performance of the motor. To evaluate the impact of detuning of this parameter, a new constant is included and defined as the ratio between estimated and real rotor time constant

$$k = \frac{\hat{c}_1}{c_1}. \tag{7}$$

As per control scheme of indirect field oriented control of induction motor, the controller equations are presented as:

$$
\begin{aligned}
&u_1 = \hat{c}_1 \frac{u_3}{u_2} \\
&u_2 = u_2^{\,0} \\
&u_3 = k_p(\omega^* - \omega) + k_i \int_0^t (\omega^*(\zeta) - \omega(\zeta))\, d\zeta
\end{aligned} \tag{8}
$$

To form closed loop equations, controller equations in (8) are included in machine system of (6) and two new state variables are defined as:

$$x_3 = \omega^* - \omega \qquad \text{and} \qquad x_4 = u_3 \tag{9}$$

Including (7)-(9) in (6), the state –variable equations of the motor for the indirect field oriented induction motor control can be expressed as:

$$
\begin{aligned}
&\dot{x}_1 = -c_1 x_1 + c_2 x_4 - \frac{k c_1}{u_2^0} x_2 x_4 \\
&\dot{x}_2 = -c_1 x_2 + c_2 u_2^0 + \frac{k c_1}{u_2^0} x_1 x_4 \\
&\dot{x}_3 = -c_3 x_3 - c_4\left[c_5(x_2 x_4 - x_1 u_2^0) - T_l - \frac{c_3}{c_4}\omega^*\right] \\
&\dot{x}_4 = (k_i - k_p c_3) x_3 - k_p c_4\left[c_5(x_2 x_4 - x_1 u_2^0) - T_l \frac{c_3}{c_4}\omega^*\right]
\end{aligned} \tag{10}
$$

where $R_r$ is rotor resistance, $L_r$ is rotor self inductance, $L_m$ is the mutual inductance, P is number of pole pairs, $\omega_{sl}$ is slip angular speed, J is inertia coefficient, $T_l$ is load torque, $\psi_{qr}$ and $\psi_{dr}$ are the quadrature and direct axis flux linkage of the rotor, $\omega$ is the rotor angular speed, $\beta$ is friction coefficient, $\omega^*$ is reference speed, $u_2^0$ is the constant reference for the rotor flux magnitude, $i_{ds}$ is the direct axis stator current and $i_{qs}$ is the quadrature axis stator current.

## III. Dynamics of the System

The dynamics of the IFOC induction motor are studied with the help of computer simulation. The simulation are carried out with the following parameter values: $c_1$=13.67, $c_2$=1.56, $c_3$=0.59, $c_4$=1176, $c_5$=2.86, $k_p$=0.001, $k_i$=0.5, $w_{ref}$=181.1 rad/s.

Fig. 1 (a) and 2(b) show the period-1 limit cycle of IFOC induction motor. Limit cycles with period-2, period-3 and period-4 are shown in Fig. 2, 3& 4 respectively. The periodicity of the limit cycle is determined by the number of loops in the phase plane or the repetition of peaks in the time plot of a state variable. The presence of infinite loops in the phase plane or non repetition of peaks in the time plot indicates chaos. The chaotic attractor is shown in Fig. 5.

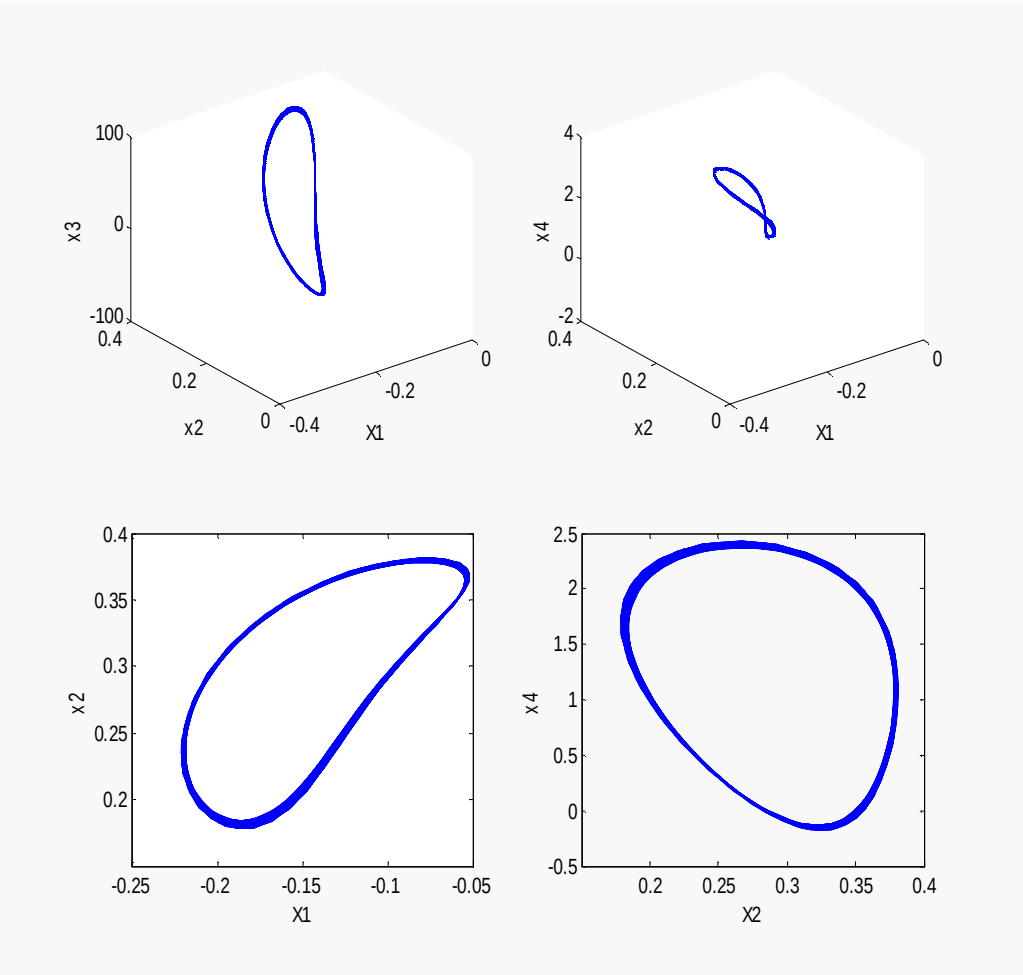

(a)

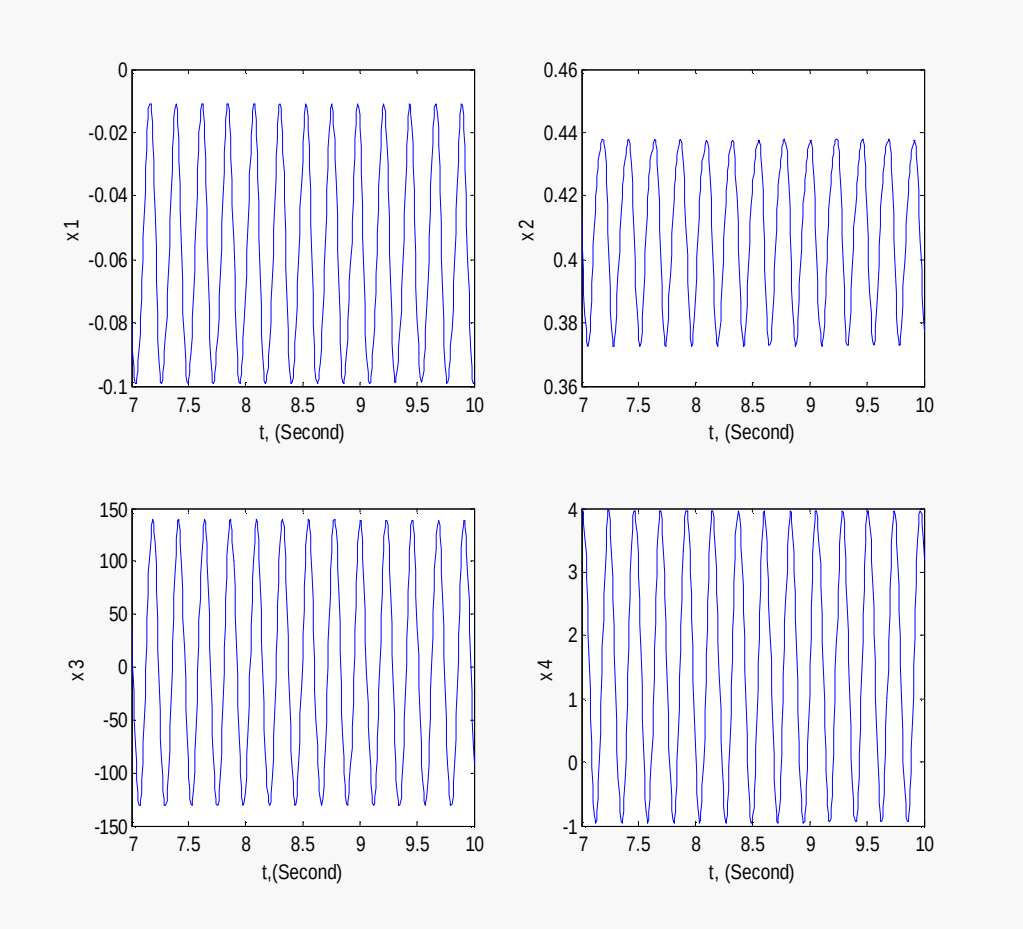

(b)

Fig: 1 (a) Phase plot at K=1.5 and Tl=2.3

(b) Time plot at K=1.5 and Tl=2.3

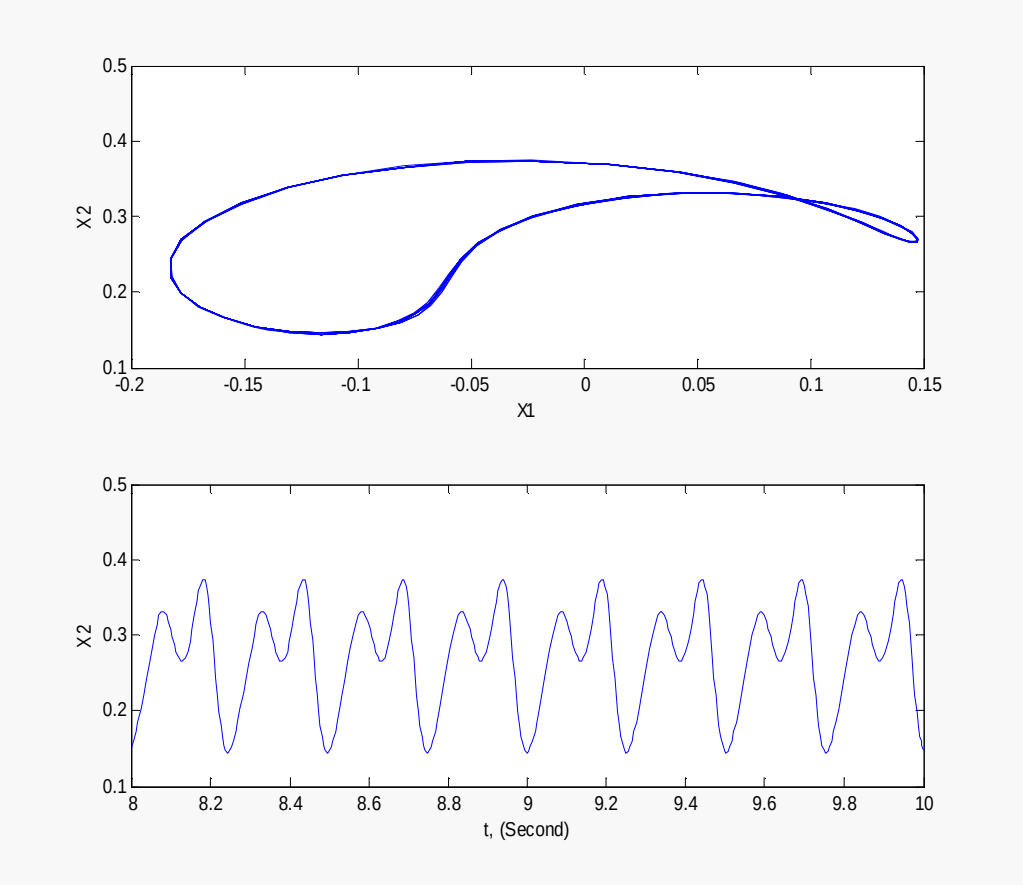

Fig: 2. Phase & Time plot at K=2.5 and Tl=0.5

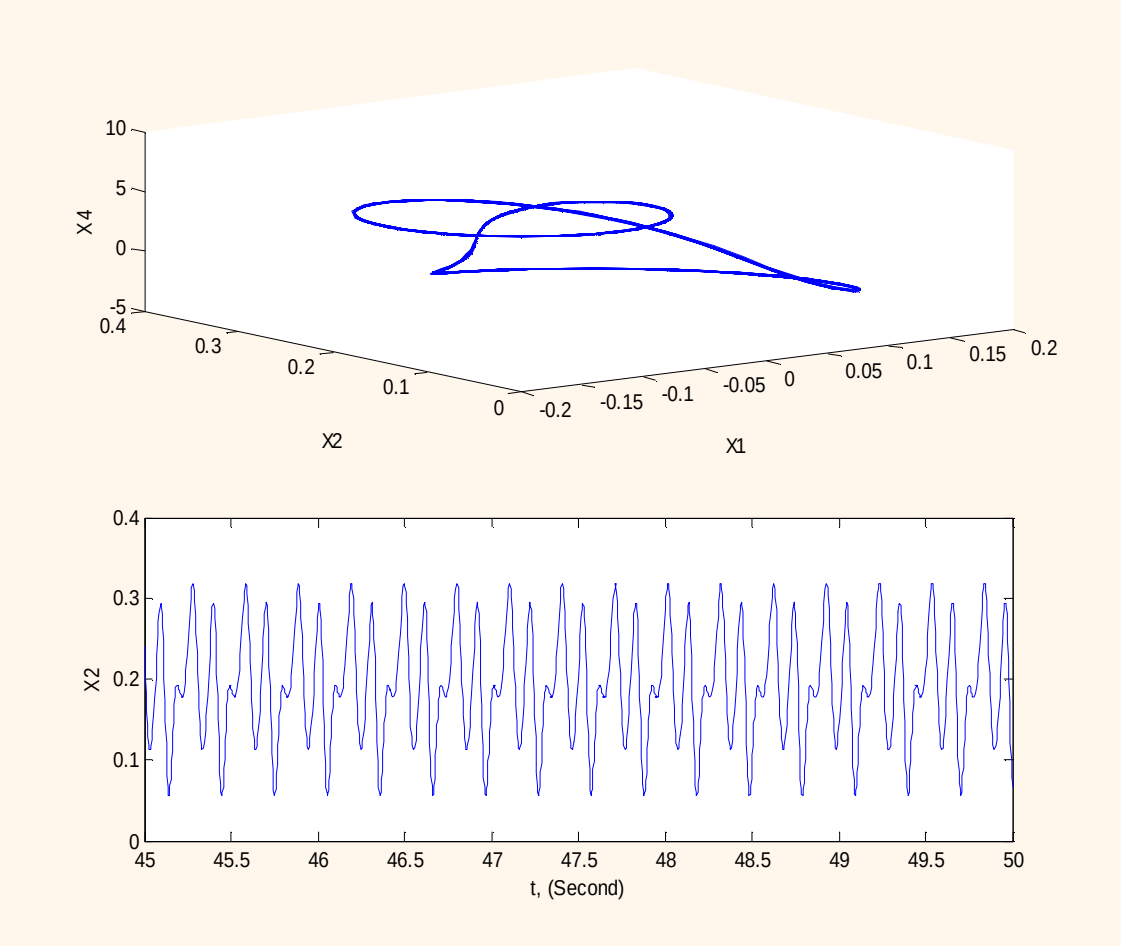

Fig. 3. Phase and time plot showing period-3 at K=3.25 and $T_l$=0.5

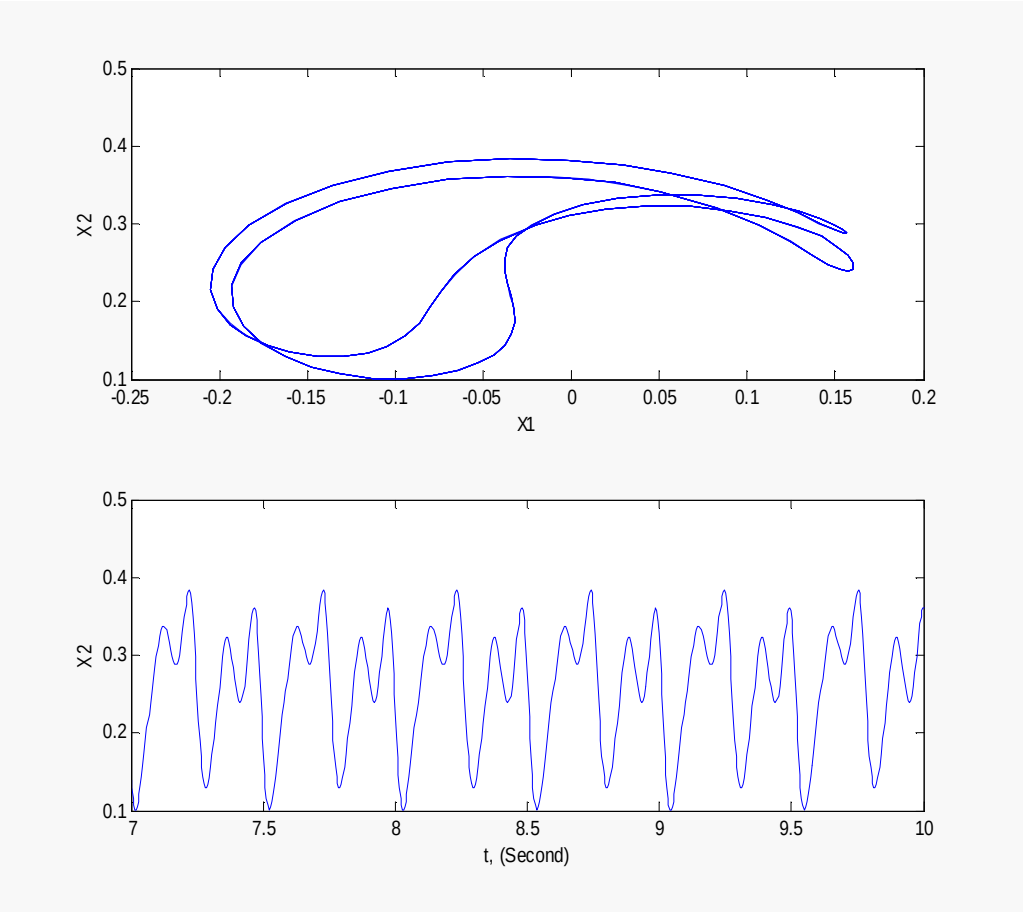

Fig. . 4. Phase & Time plot at K=2.8 and $T_l$=0.5

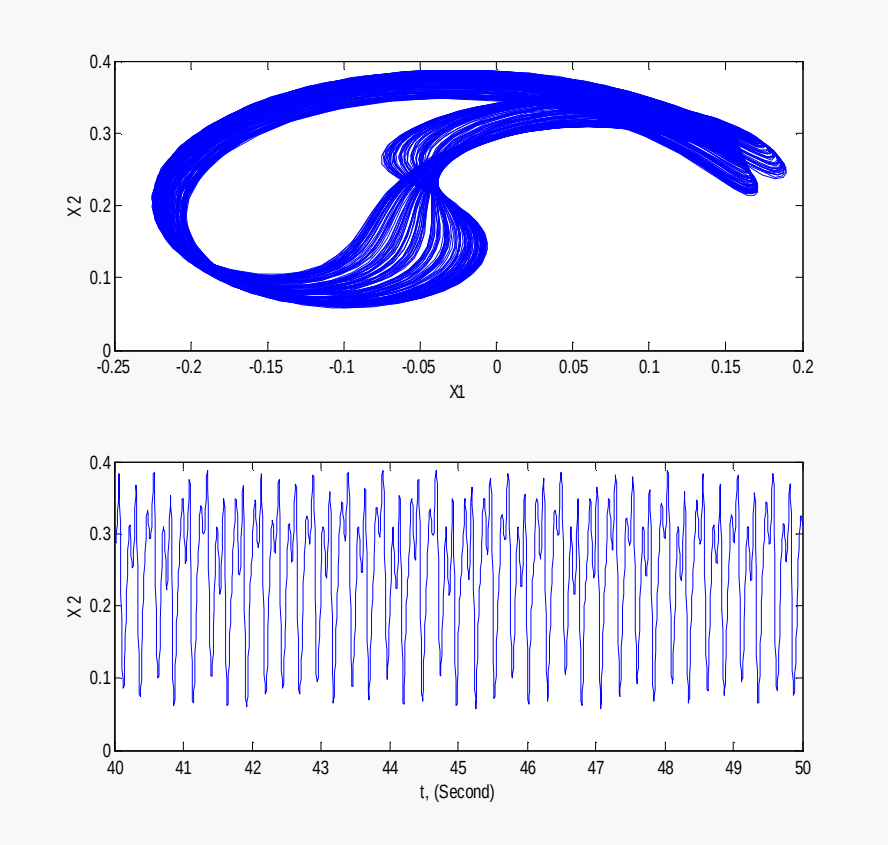

Fig:5 Phase and time plot showing Chaotic orbits at K=3.15 and $T_l$=0.5

## IV. BIFURCATION BEHAVIOR OF THE SYSTEM

A bifurcation diagram shows the long-term qualitative change (equilibria/fixed points or periodic orbits) of a system as a function of a bifurcation parameter of the system. The complete dynamics of the system with the variation of the parameter can be shown with the help of bifurcation diagram. To know the complete dynamics of the system, K (the ratio of rotor time constant to its estimate) and gain of the PI controller and load torque are taken as bifurcation parameters

### A. The Hopf bifurcation:

The Fig. 6 (a) and (b) show the phase and time plot of the IFOC induction motor at K=1.1 and load torque 0.2 Nm. It is evident from the figure that the trajectories are converging to a fixed point. The co-ordinates of the fixed point are (-0.0023, 0.4534, 0.0 0.2040). The eigenvalues of the Jacobian matrix at the fixed point are 0.5264+i27.8839, -0.5264-i27.8839, -14.2910+i0.4546, -14.2910-i0.4546.. A hopf bifurcation occurs when the value of K is increased to 1.196. A hopf bifurcation corresponds to the situation when parameter K passes through a critical value 1.196 and one pair of complex conjugate eigenvalues of the system Jacobian matrix moves from the left-half plane to right half , crossing the imaginary axis, while all other eigenvalues remain stable. At the moment of crossing, the real parts of the two eigenvalues become zero and the stability of the existing equilibrium changes from being stable to unstable. Also at the moment of crossing, a limit cycle is born. At this critical value, the eigen values are i28.2072, -i28.2072, -15.5299, -14.0929.

The eigen values of the Jacobian matrix at the corresponding equilibrium points can be computed for each value of k, giving the locus of the eigen values shown in Fig. 7. In this figure only two branches are plotted since other two branches are far away to the left and do not contribute to this bifurcation. The critical value of k can be found from the locus.

.

:

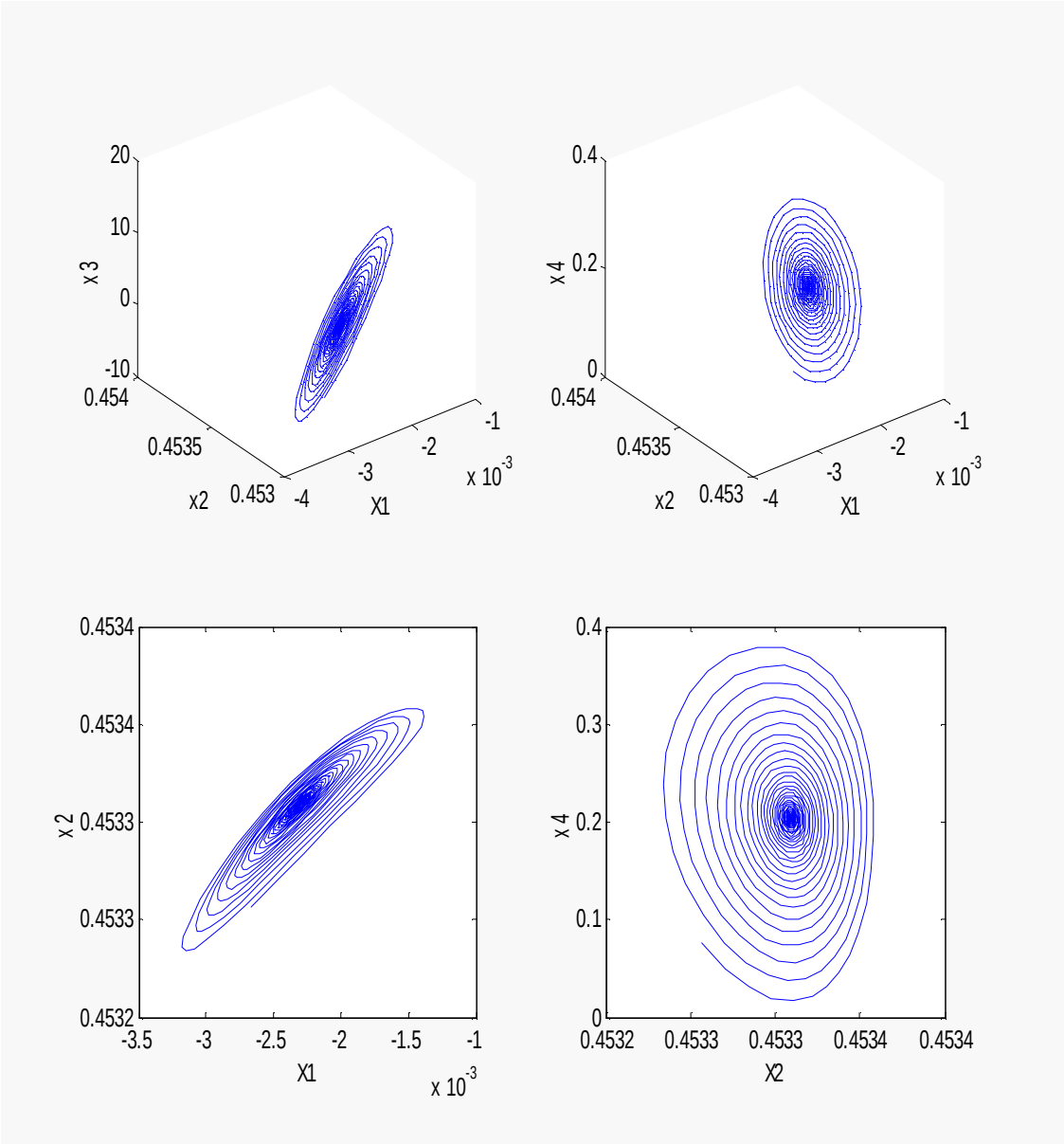

(a)

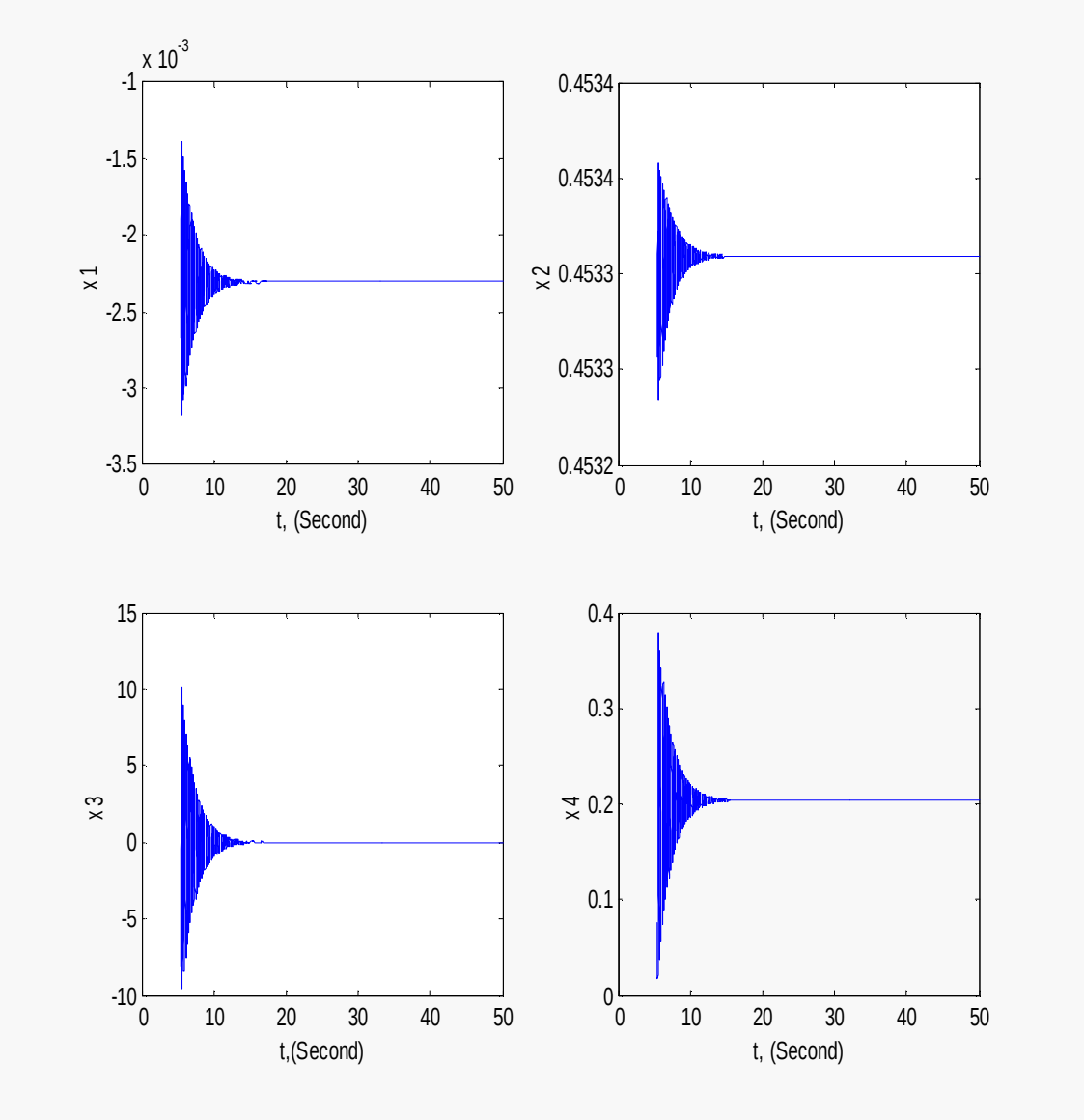

.

(b)

Fig:6. (a) Phase plot with K=1.1 and Tl=0.2

(b) Time plot with K=1.1 and Tl=0.2

### B. K as bifurcation parameter:

The bifurcation diagram shown in Fig. 8 is drawn taking K as a parameter. As K denotes the degree of tuning to the ratio of rotor time constant to its estimate, the value of K is varied from 1.5 to 5 with step size 0.001. It is evident from the bifurcation diagram that at low value of K, the state variable $x_3$ exhibits period-1 behavior. This period-1 behavior changes to period-2 sub harmonic at K=2.6. There is a transition of period-2 sub harmonic to period-4 at K=2.9.

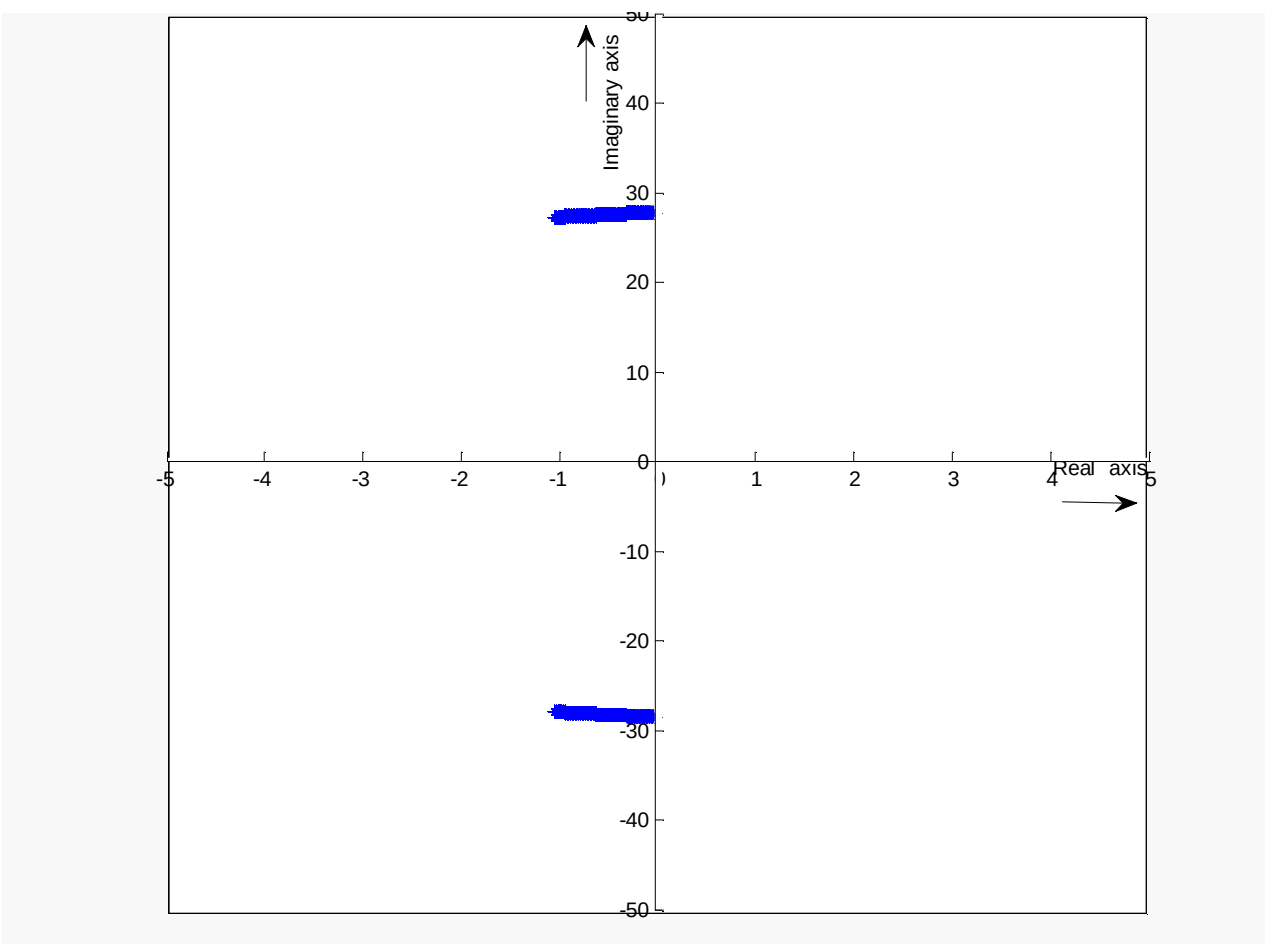

Fig. 7. Locus of the eigen values of Jacobian matrix

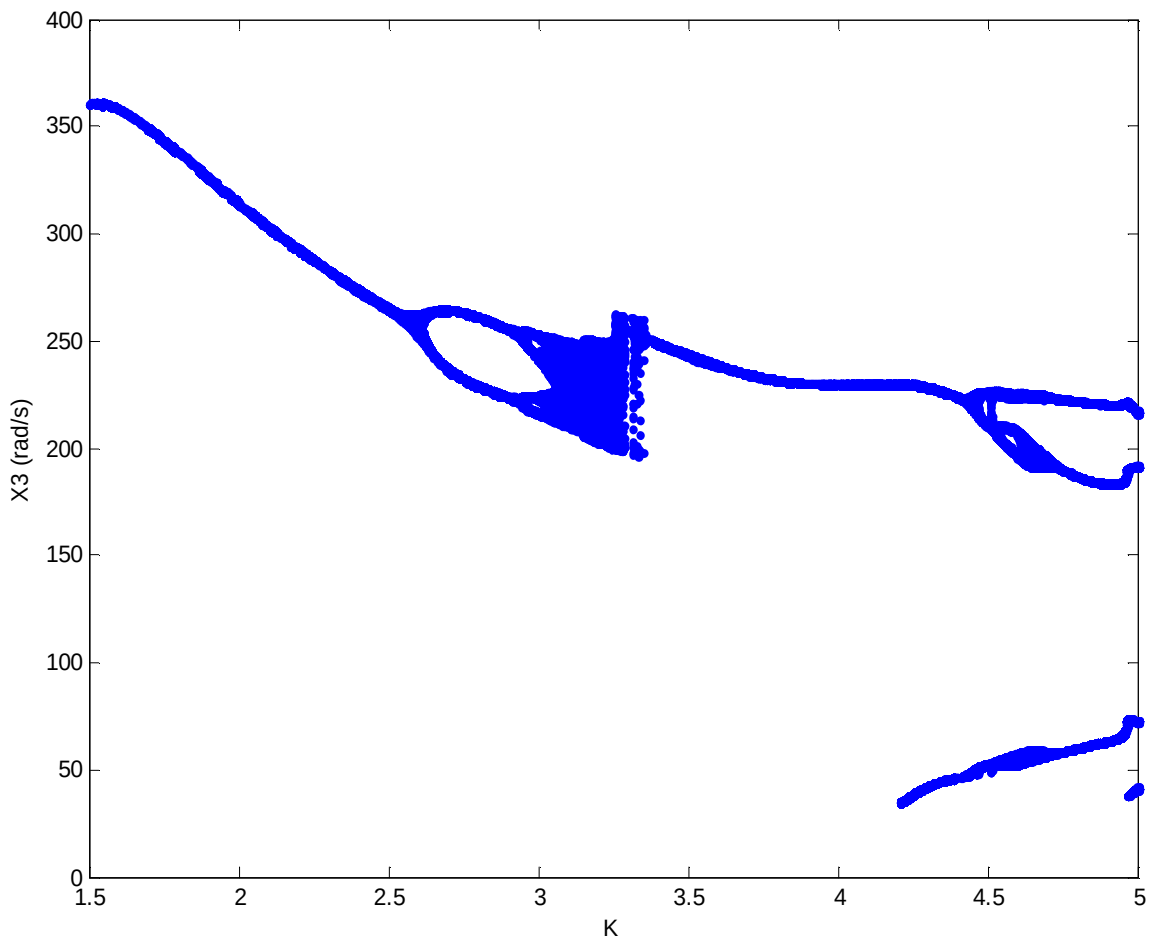

Fig.8. Bifurcation diagram with K as parameter and load torque 0.5 Nm.

The period-4 sub harmonic behavior exists for very small zone of K. Then there is a zone of chaos. In the chaotic zone, there is a very small periodic window of period-1. After the zone of chaos, there is long region of period-1 behavior. This period-1 behavior bifurcates to period-2, period-2 to period-3, then period-4. The period-4 again converges to period-3 sub harmonic at K=4.8. At K=5, period-3 bifurcates to period-4.

C. The integral gain ,$K_i$, of PI regulator as bifurcation parameter:

The bifurcation diagram is shown in Fig. 9. The gain $K_i$ is varied from 0.1 to 3. At the low value of gain, the state variable $x_3$ shows period-1 behavior and continues up to $K_i$ =0.55. Then it bifurcates to period-2 sub harmonics. It converges to period-1 again at 1.4. It is evident from the bifurcation diagram that there are patches of chaos in period-2 subharmonic region. More over there are traces of superimposed chaos in the period-2 and period-1 region of the bifurcation diagram. These behavior might be due to the existence of co-existing attractors. The period-1 behavior starting at 1.4 converges to chaos and continues up to $K_i$ =3.

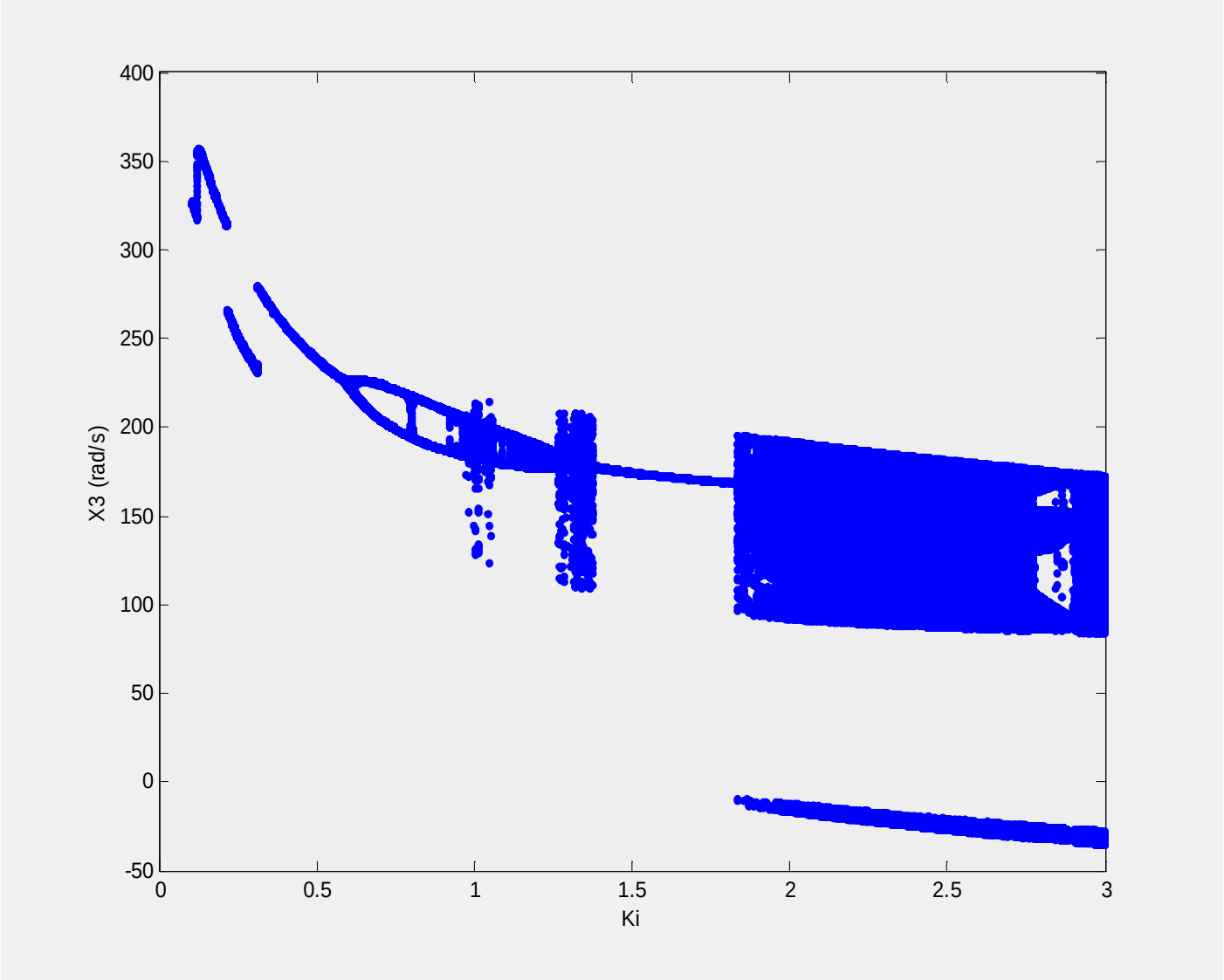

Fig: 9. Bifurcation diagram with $K_i$ as parameter and load torque 0.5 Nm.

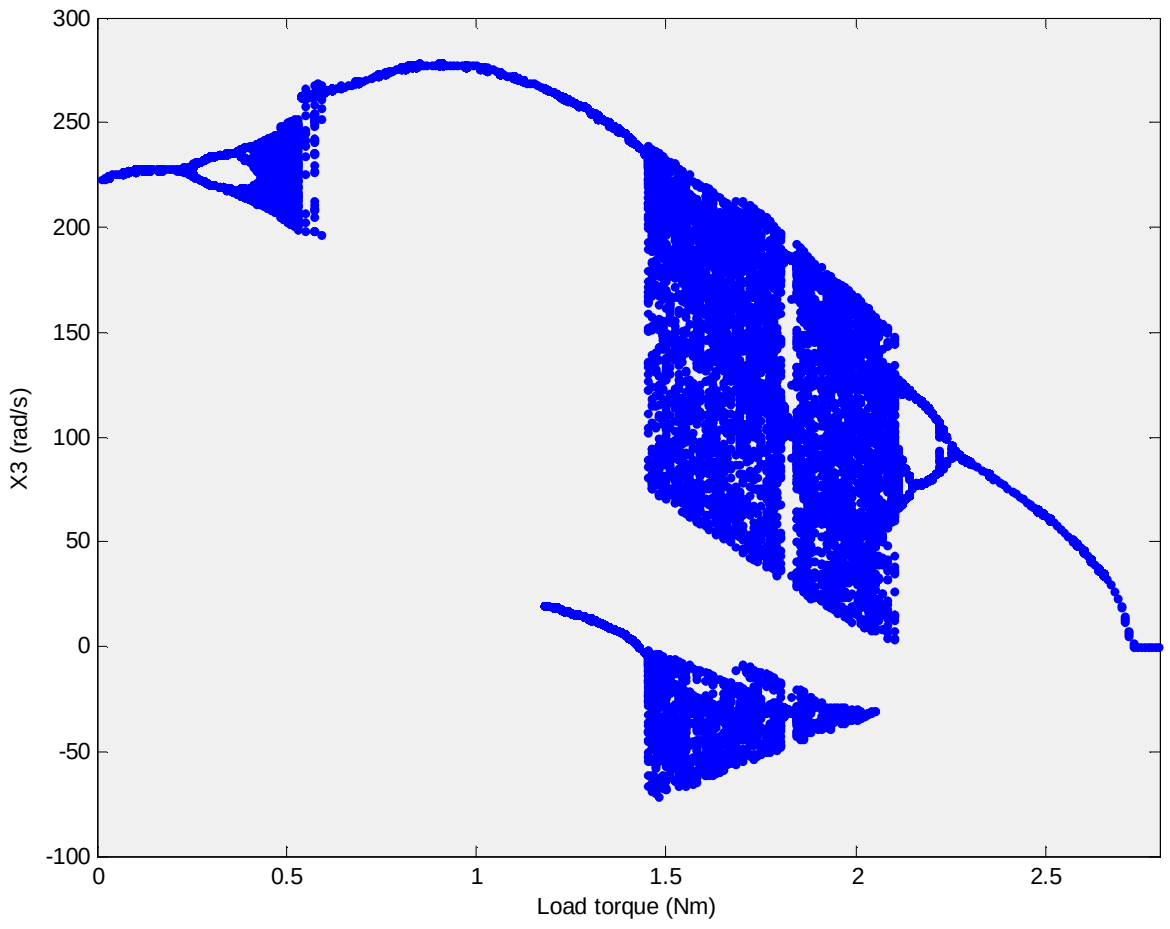

Fig.10. Bifurcation diagram with load torque as parameter.

D. The load torque, $T_l$, as parameter

The bifurcation diagram with load torque as parameter is shown in Fig. 10. The load torque is varied from 0.01 Nm to 3 Nm at a step size of 0.001. This bifurcation diagram has very interesting features. It shows identical bifurcation when the parameter is increased from 0.01Nm and decreased from 3 Nm. In both directions, period-1 bifurcates to period-2 sub harmonic, then to period-4 and then to chaos. There are periodic window embedded in the chaotic region of the bifurcation diagram obtained when parameter is decreased. Both the region of bifurcation with ascending and descending order of variation of parameter has a tail like period-1 structure connecting both the regions.

## V. . CONTROL OF CHOAS

The method of control of chaos[6] in IFOC induction motor consists of substituting a perturbation in the form in (10)

$$F(t) = K_f[x(t-\tau) - x(t)] \qquad (11)$$

Here $\tau$ is a delay time. If this time coincides with the period of ith unstable periodic orbit that to be stabilized, $T_i=\tau$, then the perturbation becomes zero for the solution of system (10) corresponding to this unstable periodic orbit $x(t)=x_i(t)$. This means that the perturbation in the form (11) does not change the solution of the system (10). The choice of appropriate value of $K_f$ will ensure the stabilization of the desired unstable periodic orbit. In the light of perturbation given in (11), the equations in (10) are modified as

$$\begin{aligned}
\dot{x}_1 &= -c_1x_1 + c_2x_4 - \frac{kc_1}{u_2^0}x_2x_4 \\
\dot{x}_2 &= -c_1x_2 + c_2u_2^0 + \frac{kc_1}{u_2^0}x_1x_4 \\
\dot{x}_3 &= -c_3x_3 - c_4\left[c_5(x_2x_4 - x_1u_2^0) - T_l - \frac{c_3}{c_4}\omega^*\right] \qquad (12) \\
\dot{x}_4 &= (k_i - k_pc_3)x_3 - k_pc_4\left[c_5(x_2x_4 - x_1u_2^0) - T_l\frac{c_3}{c_4}\omega^*\right] \\
&+ K_f[x_4(t-\tau) - x_4]
\end{aligned}$$

The bifurcation diagram shown in Fig.8 shows that the IFOC motor behaves chaotically when K=3.15 and $T_l$ =0.5 Nm. The corresponding time and phase plots are shown in Fig.5. It is intended to control period-1 unstable periodic orbit of the chaotic attractor. The period-1 orbit has been stabilized taking $\tau$=0.25 and $K_f$=10. The result is shown in Fig. 11.

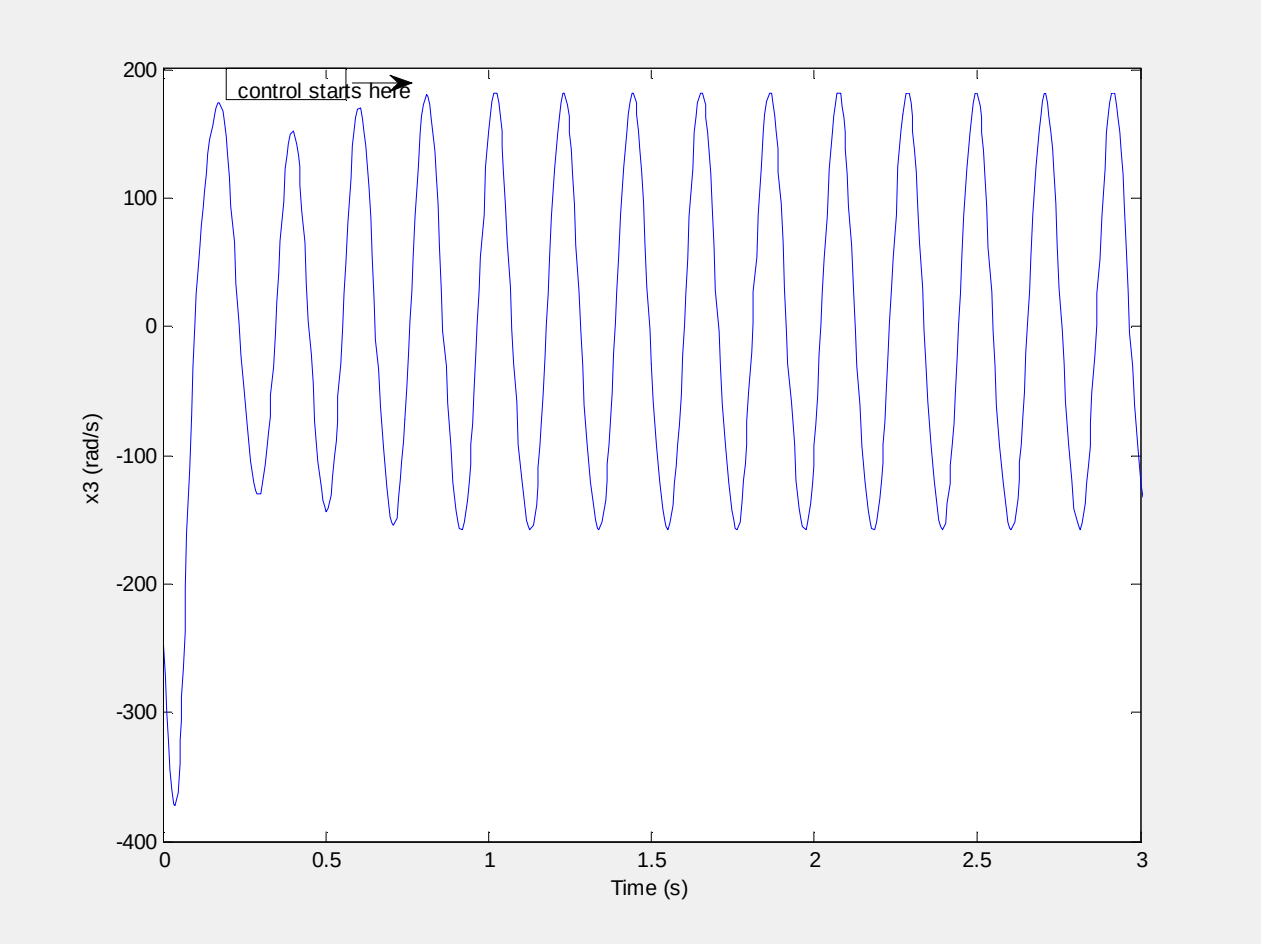

Fig. 11. Stabilized period-1 orbit of IFOC induction motor.

## VI. CONCLUSION

In order to illustrate the effects of variation of parameters on the system's performance, we have presented bifurcation analysis of IFOC induction motor. It is also shown that the IFOC motors are not only prone to instability due to Hopf bifurcation, it also exhibits limit cycles and chaos due to bifurcation other than Hopf bifurcation. This information will be very helpful for the stability analysis and for selection of appropriate parameters by design engineers to have desired performance of the motor. Moreover, there are situations where the motor behaves chaotically for certain parameter values. As for example, there may be a situation that the value of the load torque makes the speed of the motor oscillate chaotically. To ensure normal desired operation of the motor, the method of control of chaos can be applied. This method of control of chaos is very simple and can be implemented easily in experiments.